\documentclass[10pt,conference]{IEEEtran}

\usepackage[utf8]{inputenc}
\usepackage[T1]{fontenc}
\usepackage{cite}
\usepackage{amsmath,amssymb}
\usepackage{booktabs}
\usepackage{array}
\usepackage{tabularx}
\usepackage{makecell}
\usepackage{graphicx}
\usepackage{url}
\usepackage[hidelinks]{hyperref}
\usepackage{balance}

\title{Evaluating LLM-Generated Obfuscated XSS Payloads for Machine Learning-Based Detection}

\author{
\IEEEauthorblockN{Divyesh Gabbireddy}
\IEEEauthorblockA{Department of Computer Science and Engineering\\
The Pennsylvania State University\\
University Park, PA, USA\\
Email: dmg6433@psu.edu}
\and
\IEEEauthorblockN{Suman Saha}
\IEEEauthorblockA{Department of Computer Science and Engineering\\
The Pennsylvania State University\\
University Park, PA, USA\\
Email: szs339@psu.edu}
}

\begin{document}
\maketitle

\begin{abstract}
Cross-site scripting (XSS) remains a persistent web security vulnerability, especially because obfuscation can change the surface form of a malicious payload while preserving its behavior. These transformations make it difficult for traditional and machine learning-based detection systems to reliably identify attacks. Existing approaches for generating obfuscated payloads often emphasize syntactic diversity, but they do not always ensure that the generated samples remain behaviorally valid. This paper presents a structured pipeline for generating and evaluating obfuscated XSS payloads using large language models (LLMs). The pipeline combines deterministic transformation techniques with LLM-based generation and uses a browser-based runtime evaluation procedure to compare payload behavior in a controlled execution environment. This allows generated samples to be assessed through observable runtime behavior rather than syntactic similarity alone. In the evaluation, an untuned baseline language model achieves a runtime behavior match rate of 0.15, while fine-tuning on behavior-preserving source-target obfuscation pairs improves the match rate to 0.22. Although this represents a measurable improvement, the results show that current LLMs still struggle to generate obfuscations that preserve observed runtime behavior. A downstream classifier evaluation further shows that adding generated payloads does not improve detection performance in this setting, although behavior-filtered generated samples can be incorporated without materially degrading performance. Overall, the study demonstrates both the promise and the limits of applying generative models to adversarial security data generation and emphasizes the importance of runtime behavior checks in improving the quality of generated data for downstream detection systems.
\end{abstract}


\section{Introduction}
\label{sec:introduction}

Cross-site scripting (XSS) is one of the most common and persistent web security vulnerabilities. In an XSS attack, an adversary injects malicious script into a web application, and that script is then executed in a user's browser. Such attacks can lead to session hijacking, data exfiltration, page manipulation, and unauthorized actions performed on behalf of a user \cite{owaspXSS,owaspTop10}. As web applications continue to grow in complexity, detecting and preventing XSS remains an important problem for modern security systems.

A major challenge in XSS detection is obfuscation. Attackers often transform malicious payloads so that the payload appears different at the string level while preserving the same browser-visible behavior. These transformations may include character encoding, string manipulation, comment insertion, case changes, or structural rewrites. As a result, detection systems must recognize not only known payloads but also transformed variants that differ substantially in syntax.

Machine learning-based XSS detection systems are commonly trained on labeled malicious and benign inputs. These systems can improve over simple pattern matching, but their behavior is strongly affected by the distribution and quality of the training data. This issue is consistent with broader concerns raised in machine learning for security, where models trained under closed-world assumptions often fail when deployed against evolving adversarial behavior \cite{sommer2010outside}. If obfuscated payloads are underrepresented in the training data, a detector may fail to generalize to adversarial variants. If generated obfuscated samples are invalid, however, they can introduce noise and weaken the detector's connection to real attack behavior.

Large language models (LLMs) create a new opportunity for generating obfuscated XSS payloads. Transformer-based models trained on code and structured text have shown strong capability in program generation and code understanding tasks \cite{chen2021evaluating,wang2021codet5}. In principle, such models could learn complex obfuscation patterns and generate variants beyond a fixed rule set. At the same time, generation alone does not guarantee that an output preserves the runtime behavior of the original payload. This limitation is especially important for XSS, where a small change to an event handler, JavaScript function, or execution context can make a payload non-functional.

This paper asks whether LLMs can generate meaningful obfuscated XSS payloads that preserve the observed runtime behavior of their original forms. To study this question, we introduce a pipeline that combines deterministic obfuscation, LLM-based generation, and browser-based runtime evaluation. The pipeline first constructs deterministic transformation chains from original malicious payloads. Chains that preserve behavior under browser execution are converted into source-target training pairs. A language model is then fine-tuned on these behavior-filtered examples and evaluated against an untuned baseline. Finally, generated samples are assessed both for runtime behavior preservation and for their effect on downstream XSS detection.

The paper makes three contributions:
\begin{itemize}
    \item It presents a structured pipeline for constructing multi-step XSS obfuscation chains and using behavior-filtered examples for LLM training.
    \item It introduces a browser-based runtime evaluation procedure that compares generated payloads through observable execution behavior rather than syntactic similarity alone.
    \item It empirically evaluates LLM-generated obfuscations and their downstream impact on machine learning-based XSS detection, showing both the measurable benefit of fine-tuning and the remaining difficulty of behavior-preserving generation.
\end{itemize}

The results show that fine-tuning improves runtime behavior match rate from 0.15 to 0.22, a 46.7\% relative improvement over the baseline. However, most generated outputs still fail to preserve behavior, highlighting the gap between syntactic plausibility and execution-level correctness. The downstream detector evaluation shows little change in performance because the generated samples form a small portion of the overall dataset, but behavior-filtered samples can be added without materially degrading detection accuracy.

\section{Background and Related Work}
\label{sec:background}

\subsection{Cross-Site Scripting and Obfuscation}
XSS occurs when a web application includes untrusted input in generated content without sufficient sanitization, validation, or output encoding. The injected script executes in the user's browser and can access browser-level context available to the vulnerable page \cite{owaspXSS}. XSS remains a practical concern because web applications often combine user input, dynamic rendering, third-party scripts, and complex client-side behavior.

Obfuscation increases the difficulty of detection by changing the form of a payload while attempting to preserve its behavior. Common transformations include hexadecimal encoding, URL encoding, string splitting, inserted comments, whitespace manipulation, and case changes. These transformations are often combined. A detector that relies on exact strings or shallow patterns may miss a transformed payload even when the payload still triggers the same browser-visible action.

\subsection{Machine Learning for Security Detection}
Machine learning has been widely explored for security detection tasks, including intrusion detection and malicious input classification. A recurring challenge is that security data often changes over time and adversaries deliberately adapt to detectors. Sommer and Paxson argued that intrusion detection differs from many standard machine learning settings because the operational environment is not closed, the base rates are difficult, and evaluation must reflect realistic deployment assumptions \cite{sommer2010outside}. These concerns also apply to XSS detection: training data must be diverse enough to represent adversarial variation, but it must also remain valid and semantically meaningful.

Data augmentation is one way to address limited coverage of obfuscated payloads. Rule-based transformations can increase syntactic diversity, but they are limited by the predefined transformation set and may produce invalid payloads. In a security context, invalid generated samples are not harmless. If a sample looks malicious but no longer executes as an attack, it can teach a model artifacts that do not correspond to real behavior.

\subsection{LLMs for Code and Security-Relevant Generation}
LLMs trained on source code and structured text have demonstrated strong performance on code generation and code understanding tasks. Codex showed that large-scale code models can synthesize programs from natural language prompts \cite{chen2021evaluating}, while CodeT5 demonstrated that encoder-decoder models can support multiple code understanding and generation tasks \cite{wang2021codet5}. These developments motivate the use of LLMs for generating security-relevant inputs such as obfuscated XSS payloads.

However, code generation quality is not the same as runtime correctness. An LLM may produce output that appears plausible but does not execute as intended. In adversarial security domains, this problem is amplified because small syntactic changes can remove the malicious behavior. Work on vulnerability disclosure and exploitation has also shown that security-relevant behavior is shaped by dynamic, real-world conditions rather than static artifacts alone \cite{sabottke2015vulnerability}. For generated XSS payloads, this motivates an evaluation strategy that checks execution behavior directly.

This paper addresses that gap by combining LLM-based generation with browser-based runtime validation. Instead of treating syntactic similarity as sufficient, the proposed pipeline evaluates whether the generated payload produces the same observable behavior as the original under controlled execution.

\section{Methodology}
\label{sec:methodology}

This work proposes a pipeline for generating, validating, and evaluating obfuscated XSS payloads using both deterministic transformations and LLM-based generation. The goal is to determine whether generated obfuscations preserve the observable runtime behavior of the original malicious code and whether behavior-filtered samples are useful for downstream detection.

\begin{figure*}[t]
    \centering
    \input{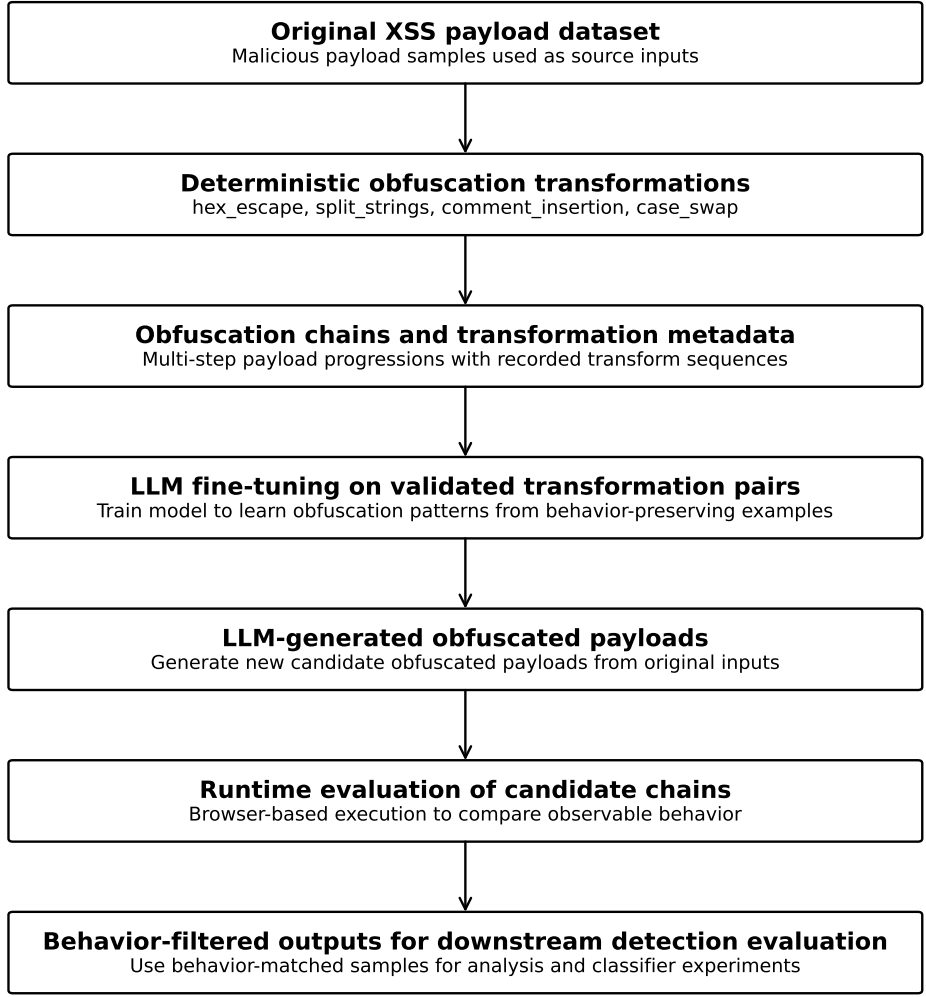}
    \caption{Overall pipeline for generating, validating, and evaluating obfuscated XSS payloads.}
    \label{fig:pipeline}
\end{figure*}

Figure~\ref{fig:pipeline} summarizes the complete process. The pipeline begins with original malicious XSS payloads. It applies deterministic obfuscation transformations to construct multi-step chains and records transformation metadata. A subset of chains is then evaluated in a controlled browser environment. Chains whose observed behavior matches the original payload are retained as behavior-preserving examples. These examples are used to fine-tune an LLM, which then generates candidate obfuscated payloads. The generated outputs are evaluated through the same runtime procedure, and behavior-filtered outputs are used in downstream classifier experiments.

\subsection{Deterministic Obfuscation Chains}
The first stage generates obfuscated variants of original malicious payloads using deterministic transformation functions. The implementation includes hexadecimal escaping, Base64 encoding with dynamic evaluation, URL encoding, string splitting, comment insertion, and case manipulation. The chain construction used in the experiments is limited to five fixed two-step recipes:
\begin{enumerate}
    \item \texttt{hex\_escape} $\rightarrow$ \texttt{split\_strings},
    \item \texttt{comment\_insertion} $\rightarrow$ \texttt{hex\_escape},
    \item \texttt{split\_strings} $\rightarrow$ \texttt{comment\_insertion},
    \item \texttt{case\_swap} $\rightarrow$ \texttt{split\_strings}, and
    \item \texttt{comment\_insertion} $\rightarrow$ \texttt{case\_swap}.
\end{enumerate}

These recipes are intended to modify the syntactic structure of a payload while attempting to preserve its browser-visible behavior. Applying multiple transformations in sequence is important because real adversarial inputs often combine more than one obfuscation strategy. For each chain, the intermediate payloads and transformation sequence are recorded as metadata.

\begin{figure}[t]
    \centering
    \input{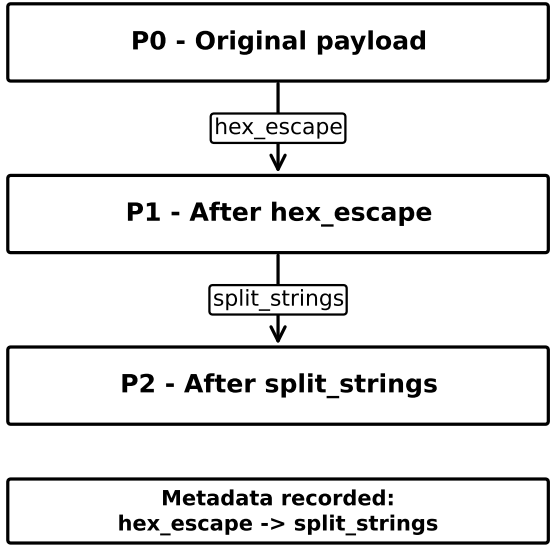}
    \caption{Example of a multi-step obfuscation chain showing sequential transformations applied to an original XSS payload.}
    \label{fig:chain}
\end{figure}

Figure~\ref{fig:chain} illustrates this idea. An original payload $P_0$ is transformed into an intermediate payload $P_1$ and then into a two-step obfuscated payload $P_2$. The fine-tuning stage uses validated $P_0 \rightarrow P_2$ pairs rather than arbitrary generated strings.

\subsection{Behavior-Filtered Training Pairs}
After constructing obfuscation chains, a subset is evaluated using the browser-based runtime checker described below. A chain is retained if the observed behavior of the final obfuscated payload matches the behavior of the original payload under the same execution conditions. Retained chains are converted into sequence-to-sequence training pairs by mapping each original payload $P_0$ to its evaluated two-step obfuscated output $P_2$.

This setup differs from traditional augmentation, which often accepts generated strings based only on transformation rules or syntactic plausibility. Here, the training set is filtered by runtime behavior before it is used for model fine-tuning. This design encourages the model to learn from examples that are not only obfuscated but also behaviorally valid under the evaluation procedure.

\subsection{LLM-Based Obfuscation Generation}
The LLM is evaluated in two configurations. The first is an untuned baseline model applied directly to the obfuscation generation task. The second is a model fine-tuned on behavior-filtered source-target pairs. Given an original payload, the model generates a candidate obfuscated payload. Unlike deterministic transformations, the model can produce variations that are not explicitly encoded in the fixed transformation recipes. However, because generation is unconstrained, not all outputs are expected to preserve runtime behavior.

\subsection{Browser-Based Runtime Evaluation}
To evaluate whether a generated payload preserves behavior, each original and generated payload is executed in a controlled browser environment. The runtime checker records observable effects, including alert triggers, console outputs, network requests, and execution errors. A generated payload is classified as behaviorally matched if its recorded execution trace matches the trace of the original payload under identical conditions.

This dynamic evaluation is more appropriate than string similarity for obfuscated XSS payloads. Two payloads may look very different but execute the same behavior, while two syntactically similar payloads may behave differently if a key execution element is changed. Runtime comparison therefore provides a stronger approximation of functional preservation than static or string-based comparison alone.

\subsection{Downstream Detection Evaluation}
Finally, generated payloads are used to evaluate their impact on machine learning-based XSS detection. Detection models are trained under three conditions: original data only, original data augmented with all generated payloads, and original data augmented only with generated payloads that pass runtime behavior matching. This comparison evaluates whether runtime-based filtering changes downstream detection performance and whether generated samples can be introduced without degrading classifier quality.

All runtime checks are conducted in a controlled experimental environment. The pipeline is intended for defensive evaluation of generated data quality and detection robustness, not for deployment of attacks against real systems.

\section{Experimental Setup}
\label{sec:setup}

This section describes the dataset, obfuscation generation process, model configuration, execution environment, and evaluation metrics used to assess LLM-generated obfuscated XSS payloads.

\subsection{Dataset}
The dataset consists of XSS payloads labeled as malicious or benign. The malicious portion is used as the source for obfuscation generation, while the benign samples are used in downstream classification experiments. This setup reflects a realistic detection scenario in which models must distinguish malicious payloads from non-malicious inputs.

\begin{table}[t]
\centering
\caption{Dataset composition and obfuscation chain generation.}
\label{tab:dataset}
\begin{tabular}{lr}
\toprule
\textbf{Dataset Property} & \textbf{Value} \\
\midrule
Total samples & 37,605 \\
Benign samples & 24,185 \\
Malicious samples & 13,420 \\
Obfuscation chains generated & 1,000 \\
Validated sample size & 200 \\
Behavior-filtered chains & 88 \\
Non-matching chains & 112 \\
Validation rate, chain-level & 44.00\% \\
\bottomrule
\end{tabular}
\end{table}

Table~\ref{tab:dataset} summarizes the dataset and validation process. From 1,000 generated chains, a random sample of 200 was evaluated using the browser-based runtime comparison procedure. Of those 200 chains, 88 were classified as behaviorally matched, yielding a chain-level validation rate of 44.00\%.

\subsection{Obfuscation Generation}
For each selected malicious payload, multiple deterministic obfuscation chains are generated by applying two-step transformation recipes. Although the implementation includes several transformation functions, the experiments use five fixed recipes involving hexadecimal escaping, string splitting, comment insertion, and case swapping. Each chain records the applied transformations and the resulting payloads. The fine-tuning stage uses the validated $P_0 \rightarrow P_2$ source-target pairs derived from these chains.

\subsection{Model Configuration}
Two model configurations are evaluated. The baseline model is used without task-specific fine-tuning. The fine-tuned model is trained on behavior-filtered source-target examples derived from validated obfuscation chains. This comparison isolates the contribution of task-specific training from the general generation capability of the model.

\begin{figure}[t]
    \centering
    \input{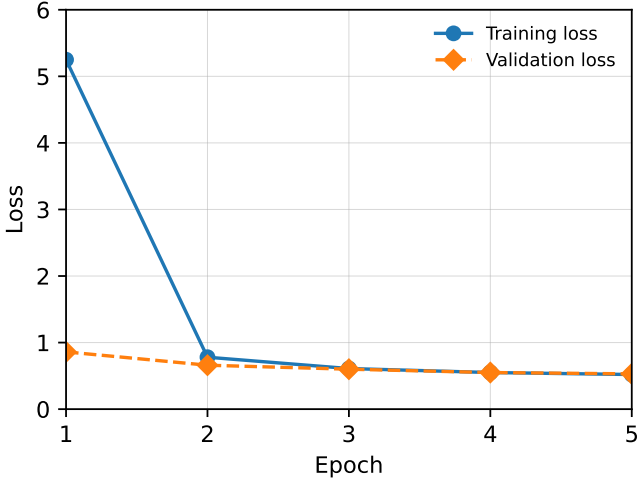}
    \caption{Training and validation loss curves across epochs for the fine-tuned language model.}
    \label{fig:loss}
\end{figure}

Figure~\ref{fig:loss} shows that both training and validation loss decrease across epochs, indicating stable learning without clear evidence of severe overfitting in the observed training run.

\subsection{Runtime Evaluation Metrics}
The primary metric is runtime behavior match rate. This is defined as the proportion of generated payloads whose observed execution trace matches the execution trace of the corresponding original payload:
\begin{equation}
    \text{Match Rate} = \frac{N_{\mathrm{matched}}}{N_{\mathrm{generated}}}.
\end{equation}
Here, $N_{\mathrm{matched}}$ is the number of behavior-matched generated payloads, and $N_{\mathrm{generated}}$ is the number of generated payloads evaluated.

This metric focuses on functional behavior rather than string-level similarity. It measures whether generated obfuscations reproduce the same browser-observed behavior as the original payloads under the controlled evaluation procedure.

\subsection{Downstream Classifier Evaluation}
For downstream detection, a Random Forest classifier with 100 estimators is trained under three data conditions: original data only, original data augmented with all generated payloads, and original data augmented with behavior-filtered generated payloads only. An 80/20 stratified split is used. Features are extracted using TF-IDF over character $n$-grams from length 3 to 5. Performance is measured using accuracy, precision, recall, and F1 score.

\section{Results}
\label{sec:results}

This section presents the runtime behavior results for LLM-generated obfuscated XSS payloads and the downstream detection results. The main question is whether generated payloads preserve the observable behavior of the original malicious input.

\subsection{Runtime Behavior Match Rate}
The evaluation compares the untuned baseline model with the fine-tuned model. For each configuration, 100 generated payloads are evaluated using the browser-based runtime comparison procedure.

\begin{table}[t]
\centering
\caption{Runtime behavior match rates of generated obfuscated payloads.}
\label{tab:model_match}
\begin{tabular}{lccc}
\toprule
\textbf{Model} & \textbf{Generated} & \textbf{Matched} & \textbf{Rate} \\
\midrule
Baseline, untuned & 100 & 15 & 0.15 \\
Fine-tuned model & 100 & 22 & 0.22 \\
\bottomrule
\end{tabular}
\end{table}

As shown in Table~\ref{tab:model_match}, the baseline model achieves a runtime behavior match rate of 0.15. This means that only 15\% of generated payloads reproduce the observed behavior of the original input under the evaluation procedure. The fine-tuned model achieves a match rate of 0.22, corresponding to a 46.7\% relative improvement over the baseline. This improvement suggests that fine-tuning on behavior-filtered obfuscation pairs helps the model learn patterns associated with behavior-preserving transformations.

At the same time, the absolute match rate remains low. Even after fine-tuning, most generated payloads fail to preserve observed runtime behavior. This result shows that LLMs can generate syntactically diverse payloads, but execution-level correctness remains a significant challenge.

\subsection{Transformation Pair Analysis}
The validation results also vary substantially across transformation pairs. Table~\ref{tab:transform_pairs} reports the number of evaluated examples, the validity rate, and the number of valid examples for each pair.

\begin{table}[t]
\centering
\caption{Runtime behavior match rates across transformation pair combinations.}
\label{tab:transform_pairs}
\begin{tabular}{lrrr}
\toprule
\textbf{Transformation Pair} & \textbf{Examples} & \textbf{Rate} & \textbf{Valid} \\
\midrule
\makecell[l]{\texttt{split\_strings} $\rightarrow$\\\texttt{comment\_insertion}} & 39 & 0.769 & 30 \\
\makecell[l]{\texttt{case\_swap} $\rightarrow$\\\texttt{split\_strings}} & 36 & 0.750 & 27 \\
\makecell[l]{\texttt{comment\_insertion} $\rightarrow$\\\texttt{case\_swap}} & 50 & 0.560 & 28 \\
\makecell[l]{\texttt{comment\_insertion} $\rightarrow$\\\texttt{hex\_escape}} & 39 & 0.051 & 2 \\
\makecell[l]{\texttt{hex\_escape} $\rightarrow$\\\texttt{split\_strings}} & 36 & 0.028 & 1 \\
\bottomrule
\end{tabular}
\end{table}

\begin{figure}[t]
    \centering
    \input{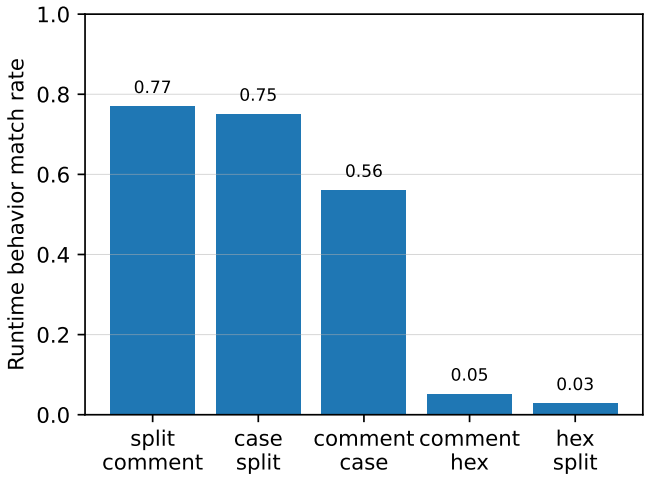}
    \caption{Runtime behavior match rates across different transformation pair combinations.}
    \label{fig:transform_rates}
\end{figure}

Figure~\ref{fig:transform_rates} visualizes the same trend. Transformation sequences involving string manipulation, such as \texttt{split\_strings} followed by \texttt{comment\_insertion}, achieve higher validity rates. Encoding-heavy sequences, especially those involving \texttt{hex\_escape}, show much lower validity. This suggests that some obfuscation strategies are more stable under transformation than others.

\subsection{Downstream Detection Performance}
To assess practical utility, a Random Forest classifier is evaluated under three training conditions. Table~\ref{tab:downstream} reports accuracy, precision, recall, and F1 score.

\begin{table}[t]
\centering
\caption{Downstream XSS detection performance under different training data conditions.}
\label{tab:downstream}
\begin{tabular}{lcccc}
\toprule
\textbf{Condition} & \textbf{Acc.} & \textbf{Prec.} & \textbf{Rec.} & \textbf{F1} \\
\midrule
Original only & 0.9987 & 1.0000 & 0.9963 & 0.9981 \\
Original + all generated & 0.9984 & 1.0000 & 0.9955 & 0.9978 \\
Original + valid only & 0.9984 & 1.0000 & 0.9955 & 0.9978 \\
\bottomrule
\end{tabular}
\end{table}

All three training conditions achieve high classification performance, with F1 scores above 0.997. The model trained on original data alone slightly outperforms the augmented conditions, but the differences are marginal. These results indicate that adding LLM-generated payloads does not improve detector performance in this setting. A likely explanation is that the generated samples represent a very small fraction of the full dataset, so the classifier is dominated by the original data distribution.

The results also show that adding generated payloads does not materially degrade performance. This is important because it suggests that behavior-filtered generated examples can be incorporated safely at small scale. The main value of runtime behavior filtering in this experiment is therefore not an immediate performance gain, but improved sample quality for future settings where generated data may form a larger share of the training set.

\section{Discussion}
\label{sec:discussion}

The results highlight both the promise and the limits of using LLMs to generate obfuscated XSS payloads. Fine-tuning improves the runtime behavior match rate from 0.15 to 0.22, but the absolute match rate remains low. This means that the model learns useful transformation patterns, yet most generated samples still fail to preserve observed behavior.

One reason for this difficulty is that XSS payloads are highly sensitive to small structural changes. A generated payload may look similar to the original but fail because it modifies an event handler, removes a function call, changes a JavaScript execution context, or alters the argument of a key operation. These errors are difficult to detect through string comparison alone. They are also difficult for an unconstrained language model to avoid because syntactic plausibility does not imply execution-level correctness.

The transformation-level results provide additional insight. String manipulation techniques tend to be more stable because they change the representation of a payload without necessarily changing the execution path. In contrast, encoding-based transformations introduce more opportunities for failure if the encoded content is not reconstructed correctly at runtime. This finding suggests that future generation pipelines may benefit from transformation-aware constraints that distinguish relatively stable obfuscations from transformations that require stricter validation.

The downstream classifier results should be interpreted carefully. The generated samples did not improve detection performance, but they also did not materially harm it. Because the augmented set is small compared with the full dataset, the classifier's behavior is largely determined by the original data distribution. In a larger augmentation setting, however, invalid generated samples could have a stronger negative effect. Runtime behavior filtering is therefore important as a data quality mechanism, especially when generated data is expected to represent a larger portion of the training set.

More broadly, the study reinforces a key distinction for security-focused LLM systems: generating code-like text is not the same as generating behaviorally correct security examples. In domains such as XSS, correctness must be evaluated in the environment where the payload executes. Browser-based runtime checking provides a practical way to filter generated examples and avoid treating non-functional outputs as valid attacks.

Several limitations remain. The runtime checker captures observable behavior in a controlled browser environment, but it may not capture all possible effects that a payload could have across browsers, application contexts, or deployment settings. The fine-tuning set is also relatively small, which may limit generalization to more complex obfuscation strategies. Finally, the downstream detector uses a conventional TF-IDF and Random Forest setup; future work should evaluate whether behavior-filtered augmentation has different effects for neural classifiers or detectors trained specifically for robust adversarial generalization.

\section{Conclusion}
\label{sec:conclusion}

This paper presented a structured pipeline for generating and evaluating obfuscated XSS payloads using LLMs. The pipeline combines deterministic transformation chains, behavior-filtered fine-tuning data, browser-based runtime validation, and downstream classifier evaluation. By evaluating generated payloads through observable browser behavior rather than syntactic similarity alone, the approach provides a stronger basis for assessing whether generated obfuscations remain meaningful security examples.

The experiments show that fine-tuning on behavior-filtered source-target pairs improves runtime behavior match rate from 0.15 to 0.22. This indicates that task-specific training helps the model learn behavior-preserving obfuscation patterns. However, the low absolute match rate shows that current LLMs still struggle to preserve execution behavior reliably. The downstream classifier evaluation further shows that generated samples do not improve detection performance in this setting, although behavior-filtered samples can be added without materially degrading performance.

These findings suggest that LLM-generated security data should be paired with runtime validation before being used in detection pipelines. Future work can explore larger fine-tuning sets, constrained decoding, transformation-aware generation, multi-browser validation, and hybrid rule-based plus generative approaches that better balance diversity and correctness.

\section*{Acknowledgment}
This manuscript is adapted from an undergraduate honors thesis completed at the Schreyer Honors College, The Pennsylvania State University. The authors thank Mohamed Almekkawy for serving as the thesis honors adviser and for his support during the thesis process.

\balance
\bibliographystyle{IEEEtran}
\bibliography{references}

\end{document}